\documentclass[aip,twocolumn,groupedaddress]{revtex4}
\usepackage{graphicx}
\usepackage{bm}
\usepackage{times}
\usepackage{amsmath}
\usepackage{float}

\begin{document} 
 
\title{X-ray scattering from a plane parallel slab using radiation transfer considerations\footnote{Submitted to J. Appl. Cryst.}}

\author{Stanislav Stoupin}
\email{sstoupin@cornell.edu} 
\affiliation{Cornell High Energy Synchrotron Source, Cornell University, Ithaca, NY 14853, USA}

\begin{abstract} 
X-ray scattering power of a plane parallel homogeneous slab of material is derived using radiation intensity transfer equations. 
The scattering power scales with the ratio of the scattering coefficient of interest to the total attenuation coefficient. 
The results can be used to guide the choice of slab thickness, scattering geometry and photon energy to maximize scattering power 
in both elastic and inelastic X-ray scattering experiments.  
\end{abstract} 

\maketitle
%%%%%%%%%%%%%%%%%%%%%%%%%%%%%%%%%%%%%%%%%%%%%%%%%%%%%%%%%%%%%%%%%%%%%%%%%%%%%%%%%%%%%%%%%%%%%%%%%%%%%%%%%%%%%%%%%%%%%%%%%%%%%%%%%%%%%%%%%%%%%%%%%%%%%%%%%%%%%%%%%%%%%%%%%%%%%%%%%%%%%
\section{Introduction}
\begin{figure}
\includegraphics[width=0.5\textwidth]{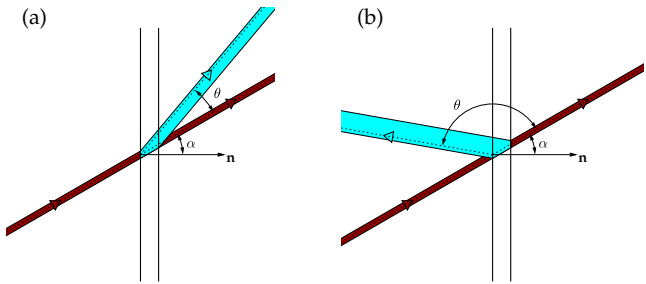}
\caption{Scattering from the parallel slab in transmission geometry (a) and in reflection geometry (b).}
\label{fig:1}
\end{figure}

%%%%%%%%%%%%%%%%%%%%%%%%%%%%%%%%%%%%%%%%%%%%%%%%%%%%%%%%%%%%%%%%%%%%%%%%%%%%%%%%%%%%%%%%%%%%%%%%%%%%%%%%%%%%%%%%%%%%%%%%%%%%%%%%%%%%%%%%%%%%%%%%%%%%%%%%%%%%%%%%%%%%%%%%%%%%%%%%%%%%%
The optimal choice of photon energy and sample thickness in X-ray scattering experiments deserves clarifications based on clear understanding of radiation transfer 
in specimens of finite thickness. 
The well-known practical requirement ("rule of thumb") for optimization of scattered intensity in transmission through a sample states that 
the optimal sample thickness should be approximately equal to the radiation attenuation length $\Lambda$:
\begin{equation}
T_{opt} \simeq \gamma_0/\mu,
\label{eq:Topt00}
\end{equation}
where $\mu = 1/\Lambda$ is the total attenuation coefficient \footnote{The total attenuation coefficient is often termed as total absorption coefficient. However, it should not be confused with the photoelectric absorption coefficient, since attenuation due to X-ray scattering from atoms with low atomic number can prevail at sufficiently high photon energies.} and $\gamma_0$ is the direction cosine for the X-ray beam incident on the sample (the angle between the wavevector of the incident beam and the inward normal to the sample surface). Intuitively it is clear that to maximize scattering in reflection from the sample its thickness should be $T \gtrsim T_{opt}$.
The basic sample geometry is that of the infinite parallel slab of material as shown in Fig.\ref{fig:1}. 
The direction cosine of the incident beam $\gamma_0 = \cos{\alpha}$ and the direction cosine of the scattered beam is  $\gamma_s = \cos{(\alpha+\theta)}$, where $\alpha$ is the incidence angle and $\theta$ is the scattering angle. By definition, $\gamma_0 > 0$, $\gamma_s > 0$ for the transmission geometry and $\gamma_s < 0$ for the reflection geometry.

Some textbooks provide additional relevant details \cite{Cullity_book, Schuelke_book, Stribeck_book}. For example, \cite{Schuelke_book} includes expressions for the scattering power (fraction of photons scattered) from a parallel slab in the symmetric ($\gamma_0$ = $|\gamma_s|$)) transmission and reflection geometries. 
A general recipe for the calculation of the absorption correction (here termed as attenuation correction) is given by International Tables for Crystallography \cite{Maslen-ITC-C}. 
\begin{equation}
A = \frac{1}{V}\int{\exp{(-\mu\Gamma)}dV},
\label{eq:A}
\end{equation}
where the integral is taken over the sample volume $V$ exposed to X-rays and $\Gamma$ is the path length of the X-ray beam in the sample (the sum of the path lengths for the penetrating incident beam and the scattered beam). Several expressions for special cases of the parallel slab geometry are tabulated \cite{Maslen-ITC-C}. Meanwhile, the absorption correction for the parallel slab in case of arbitrary angle of incidence and arbitrary scattering angle can be readily derived using Eq.~(\ref{eq:A}) (e.g., \cite{Cullity_book}).

In this work it is shown that consideration of intensity radiation transfer (Darwin-Hamilton equations) in the plane parallel slab while neglecting effects of multiple scattering leads to the same expressions as the ones derived using Eq.~(\ref{eq:A}). The solutions are expressed as functions of the momentum transfer for X-rays. The dependence of the scattering power on the momentum transfer due to the parallel slab geometry is analyzed and illustrated. Finally, energy dependence of the leading term, the scattering-attenuation contrast (the ratio of the scattering coefficient of interest $\sigma$ to the attenuation coefficient) is illustrated assuming isotropic (average) cross sections, which can guide the choice of optimal photon energy. While the choice of photon energy and the slab thickness can maximize the contrast and thus optimize the scattering power, the contrast has to remain small ($\sigma_0/\mu \ll 1$) to avoid multiple scattering effects. 

\section{Attenuation correction}
The path of the X-ray beam in the parallel slab is
\begin{eqnarray}
\Gamma &=& 
\begin{cases}
    \frac{z}{\gamma_0} + \frac{(T-z)}{\gamma_s}, & \text{transmission} \ (\gamma_s>0) 
    \\
    \frac{z}{\gamma_0} - \frac{z}{\gamma_s}, & \text{reflection} \ (\gamma_s<0),
\end{cases}
\label{eq:path}
\end{eqnarray}
where $T$ is the slab thickness and $z$ is the variable depth at which scattering event occurs. 
For the photon energies of practical importance in X-ray scattering experiments (e.g., $E \lesssim$~100~keV) the non-negligible attenuation processes are photoabsorption, elastic (Rayleigh) and inelastic (Compton) scattering. 
A general case of scattering can be considered where the scattered photon can loose a fraction of its energy (i.e., inelastic scattering). The scattered beam of interest, propagating in a direction defined by $\gamma_s$ will encounter attenuation coefficient $\mu_s(E_s)$, which can be different from that of the incident beam $\mu_0(E_0)$. 
Using Eqs. \ref{eq:A} and \ref{eq:path} for the parallel slab one can readily obtain for the scattering power:

\begin{eqnarray}
\begin{split}
s & = \sigma_0 \frac{T}{\gamma_0} A \\ 
  & =
\begin{cases}
    \frac{\sigma_0}{\mu_s b - \mu_0} e^{-\mu_0 T/\gamma_0} \Big[1 - e^{-(\mu_s b - \mu_0) T/\gamma_0} \Big], & \gamma_s>0,
    \\
    \frac{\sigma_0}{\mu_0 - \mu_s b} \Big[1 - e^{-(\mu_0 - \mu_s b) T/\gamma_0} \Big], & \gamma_s<0,
\end{cases}
\end{split}
\label{eq:solution}
\end{eqnarray}
where $b = \gamma_0/\gamma_s$ is the asymmetry ratio of the scattering geometry and $\sigma_0 = \sigma_0^s(E_0)$ is the scattering coefficient (per unit thickness, similarly to $\mu$) of the incident beam with photon energy $E_0$ into the scattering channel with photon energy $E_s$. 

%Equation (\ref{eq:A}) can be generalized as attenuation correction, if $\mu_a$ is replaced with $\mu$, the attenuation coefficient.  

The attenuation correction does not take into account effects of multiple scattering, which can be split into two categories. If the slab is a single crystal in the Bragg diffraction condition (elastic scattering), multiply scattered photons can add up coherently such that the resulting scattering direction is well defined (to the extent defined by the divergence/bandwidth of the incident beam and the perfection of the crystal structure). A detector placed to capture photons scattered in this direction will probe the corresponding momentum transfer. 
On the other hand, for any slab of material the other type of multiple scattering exists where the photons traversing the slab exhibit a few different momentum transfers (two as the strongest contribution) prior to arriving at the detector. For these photons the directional placement of the detector does not correspond to a particular momentum transfer. Therefore, multiple scattering of this second type should be avoided in X-ray scattering experiments as long as resolving the momentum transfer remains of primary interest. Dedicated studies show that such multiple scattering (both elastic and inelastic) is non-negligible at photon energies of $E \gtrsim$~30~keV  for materials of low atomic numbers (e.g., \cite{Warren66, Fajardo98}). 

%which, unlike those for neutrons \cite{Sears75}, are often assumed to be negligible for X-rays.  
%However, at photon energies of $E \gtrsim$~30~keV for a scattering medium of low atomic numbers the scattering cross sections become comparable to or greater than the photoabsorption cross section.
%%%%%%%%%%%%%%%%%%%%%%%%%%%%%%%%%%%%%%%%%%%%%%%%%%%%%%%%%%%%%%%%%%%%%%%%%%%%%%%%%%%%%%%%%%%%%%%%%%%%%%%%%%%%%%%%%%%%%%%%%%%%%%%%%%%%%%%%%%%%%%%%%%%%%%%%%%%%%%
\section{Simplified radiation transfer equations and their solutions}\label{sec:theory}
%%%%%%%%%%%%%%%%%%%%%%%%%%%%%%%%%%%%%%%%%%%%%%%%%%%%%%%%%%%%%%%%%%%%%%%%%%%%%%%%%%%%%%%%%%%%%%%%%%%%%%%%%%%%%%%%%%%%%%%%%%%%%%%%%%%%%%%%%%%%%%%%%%%%%%%%%%%%%%
%\subsection{Problem formulation}

Darwin-Hamilton equations \cite{Dietrich65,Sears97-1,Sabine-ITC-C} describe radiation intensity transfer in X-ray diffraction from a mosaic crystal.
In the Bragg diffraction condition where multiple scattering of the first type is expected to be the dominant scattering mechanism, 
the contribution of the second type of multiple-scattering is neglected. 
The notion of a crystal structure is specific to the formulation of the scattering coefficient $\sigma$. Therefore, assuming that the second type of multiple scattering is avoided and that wave mixing/interference effects are negligible the Darwin-Hamilton equations can be generalized to describe radiation transfer in X-ray scattering experiments. The generalized equations are:

\begin{eqnarray}
\frac{dI_0(E_0)}{dt} = - \frac{\mu(E_0)}{\gamma_0} I_0 (E_0) + \frac{\sigma_s^0(E_s)}{|\gamma_s|} I_s(E_s), \label{eq:DH1} \\
\frac{dI_s(E_s)}{dt} = - \frac{\mu(E_s)}{\gamma_s} I_s(E_s) \pm \frac{\sigma_0^s(E_0)}{\gamma_0} I_0(E_0)  ,
\label{eq:DH2}
\end{eqnarray}
where $I_0$ and $I_s$ are the intensities of the incident and scattered beams respectively, $t$ is the coordinate along the inward normal to the slab surface, $\gamma_0$ and $\gamma_s$ are the direction cosines for the transmitted and scattered beams respectively. The upper sign corresponds to the beam scattered in transmission through the slab and the lower sign corresponds to the beam scattered in the reflection geometry.  

%$\sigma (E_s)$ is the scattering coefficient of the process of interest,
 
Unlike the original Darwin-Hamilton equations, which describe the elastic process, scattering 
can occur at a different photon energy $E_s$ (i.e., inelastic scattering process). The scattering coefficient of interest is $\sigma_0^s (E_0)$ corresponds to scattering of the primary beam with photon energy $E_0$ into the scattering channel with a photon energy $E_s$. Similarly, re-scattering of the scattered beam back into the primary beam is described with $\sigma_s^0 (E_s)$. In general, $\mu$,~$\sigma$~and~$I_s$ are functions of the direction of propagation, which is omitted from the notation for brevity.    
Aside from the case of Bragg diffraction in crystalline samples a common situation in X-ray scattering experiments is $I_s << I_0$, while the scattering power of interest is relatively weak 
$\sigma_0^s \approx \sigma_s^0 << \mu$. In these conditions, the last term in Eq.~\ref{eq:DH1}, $\frac{\sigma_s^0(E_s)}{|\gamma_s|} I_s(E_s)$ is a second order quantity (aside from cases of extreme grazing emergence $|\gamma_s| \simeq 0$ not considered here). In other words, an assumption is made that the multiple scattering of the first type is small. 
The first equation resolves as the well known Beer-Lambert law
\begin{equation}
I_0(E_0) = I_0^0 \exp{\big[ -\mu (E_0) t /\gamma_0 \big]},
\label{eq:BL}
\end{equation}
where $I_0^0$ is the intensity of the incident beam at the entrance surface. 
Thus the radiation transfer equations become decoupled. Upon substitution of Eq. (\ref{eq:BL}) the equation for the scattered intensity is
\begin{equation}
\frac{dI_s}{dt} = - \frac{\mu_s}{\gamma_s} I_s \pm \frac{\sigma_0 I_0^0}{\gamma_0} \exp(-\mu_0 t /\gamma_0). 
\label{eq:Is}
\end{equation}
Here, for brevity $I_s = I_s(E_s)$, $\sigma_0 = \sigma_0^s(E_0)$, $\mu_s = \mu(E_s)$ and $\mu_0 = \mu(E_0)$.

Equation (\ref{eq:Is}) can be easily solved analytically using appropriate boundary conditions.
For the transmission case $I_s|_{t=0}$~=~0 and the scattering power $s = I_s|_{t=T}/I_0^0$ is
found for the slab of thickness $T$. 
For the reflection case $I_s|_{t=T}$~=~0 and the scattering power $s = I_s|_{t=0}/I_0^0$ is found at the entrance surface. 
It is easy to see that the resulting expressions are the same as those derived in the previous section (Eq.~(\ref{eq:solution})). 

\section{Analysis of solutions, special cases} 

%\begin{equation}
%s = \frac{\sigma_0}{\mu_s b - \mu_0} e^{-\mu_0 T/\gamma_0} \Big[1 - e^{-(\mu_s b - \mu_0) T/\gamma_0} \Big]
%\label{eq:Tsol0}
%\end{equation}
\subsection{Transmission geometry}

In the transmission geometry an optimal thickness exists where the scattered intensity attains its maximum value. 
The optimal thickness and the scattering power for this thickness are
\begin{eqnarray}
T_{opt} = \frac{\gamma_0 \ln [b \mu_s / \mu_0 ]}{\mu_s b - \mu_0}, \label{eq:Topt0_T} \\ 
s_{opt} = \frac{\sigma_0}{b \mu_s} \exp\bigg[- \frac{\mu_0 \ln [b\mu_s/\mu_0] }{ \mu_s b - \mu_0 } \bigg]. \label{eq:Topt0_s}
\end{eqnarray}

For experiments where the elastically scattered component is detected (e.g., small angle or wide angle X-ray scattering (SAXS/WAXS))  
the above expressions can be simplified by letting $\mu_0 = \mu_s = \mu$. Also, the subscript is 
omitted in the notation for the scattering coefficient $\sigma_0 = \sigma$.
The scattering power is
\begin{equation}
s = \frac{\sigma}{(b-1)\mu} e^{-\mu T/\gamma_0} \Big[1 - e^{-\mu (b - 1) T/\gamma_0} \Big]. 
\label{eq:Tsol1}
\end{equation}

Eqs. \ref{eq:Topt0_T} and \ref{eq:Topt0_s} are reduced to: 

\begin{eqnarray}
T_{opt} = \frac{\gamma_0 \ln b}{(b - 1)\mu}, \label{eq:Topt1_T} \\
s_{opt} = \frac{\sigma}{b \mu} \exp\bigg[- \frac{\ln b }{b - 1} \bigg]. \label{eq:Topt1_s}
\end{eqnarray}

These simplified expressions (Eqs. \ref{eq:Tsol1}, \ref{eq:Topt1_T} and \ref{eq:Topt1_s}) are also valid for nonresonant inelastic experiments, where $\mu(E)$ is a smooth function and 
the scattered intensity is measured selectively using an energy discriminating detector (including use of a crystal analyzer) at a reasonably small energy offset, such that $\mu_s \simeq \mu_0 = \mu$. 
For the symmetric scattering geometry ($b = 1$), and in the regime of small angle scattering ($\gamma_0 \simeq \gamma_s$) Eqs.~\ref{eq:Tsol1},~\ref{eq:Topt1_T}~and~\ref{eq:Topt1_s} are reduced to the well known expressions:

\begin{equation}
s = \frac{\sigma T}{\gamma_0} e^{-\mu T/\gamma_0} ,
\label{eq:Tsol2}
\end{equation}

\begin{eqnarray}
T_{opt} = \frac{\gamma_0}{\mu}, \label{eq:Topt2_T}\\  
s_{opt} = \frac{\sigma}{\mu} e^{-1}. \label{eq:Topt2_s}
\end{eqnarray}
%%%%%%%%%%%%%%%%%%%%%%%%%%%%%%%%%%%%%%%%%%%%%%%%%%%%%%%%%%%%%%%%%%%%%%%%%%%%%%%%%%%%%%%%%%%%%%%%%%%%%%%%%%%%%%%%%%%%%%%%%%%%%%
\subsection{Reflection geometry}

%\begin{equation}
%s = \frac{\sigma_0}{\mu_0 - \mu_s b} \Big[1 - e^{-(\mu_0 - \mu_s b) T/\gamma_0} \Big]
%\label{eq:Rsol0}
%\end{equation}

For the elastic and nonresonant inelastic experiments as described in the previous subsection ($\mu_0 = \mu_s = \mu$, $\sigma_0 = \sigma$):

\begin{equation}
s = \frac{\sigma}{(1 - b)\mu} \Big[1 - e^{-\mu(1 - b) T/\gamma_0} \Big],
\label{eq:Rsol1}
\end{equation}

which is reduced to 

\begin{equation}
s = \frac{\sigma}{2\mu} \Big[1 - e^{-2\mu T/\gamma_0} \Big],
\label{eq:Rsol2}
\end{equation}
for the symmetric scattering geometry ($b = -1$).
From Eqs.~(\ref{eq:Tsol2})~and~(\ref{eq:Rsol2}) it follows that in the case of small attenuation 
$\mu T/\gamma_0 \ll 1$ the scattering power is $s \simeq \frac{\sigma T}{\gamma_0}$, i. e. proportional to the effective slab 
thickness $T/\gamma_0$ in both transmission and reflection cases. Otherwise, it is proportional to the ratio $\sigma/\mu$, which will be further referred
to as scattering-attenuation contrast.

%%%%%%%%%%%%%%%%%%%%%%%%%%%%%%%%%%%%%%%%%%%%%%%%%%%%%%%%%%%%%%%%%%%%%%%%%%%%%%%%%%%%%%%%%%%%%%%%%%%%%%%%%%%%%%%%%%%%%%%%%%%%%%%%%%%%%%%%%%%%%%%%%%%%%%%%%%
\section{Dependence on the momentum transfer for X-rays}
%%%%%%%%%%%%%%%%%%%%%%%%%%%%%%%%%%%%%%%%%%%%%%%%%%%%%%%%%%%%%%%%%%%%%%%%%%%%%%%%%%%%%%%%%%%%%%%%%%%%%%%%%%%%%%%%%%%%%%%%%%%%%%%%%%%%%%%%%%%%%%%%%%%%%%%%%%%%
For X-rays the scattering angle $\theta$ is conveniently related to the modulus of momentum transfer $Q = 2 k_0 \sin(\theta/2)$.
The change in the modulus of the wavevector $k_0$ upon scattering is either zero (elastic scattering) or very small (quasielastic scattering).  Assuming normal incidence of X-rays ($\gamma_0$=1) the exit angle for the scattered beam with respect to the normal becomes equal to the scattering angle. 
The direction cosine of the scattered beam is
\begin{equation}
\gamma_s = 1 - 2q^2, 
\label{eq:q}
\end{equation}
where $q = \frac{Q}{2k_0} = \sin(\theta/2)$ is the reduced momentum transfer. 
In the transmission geometry $0 \leq q < 1.0/\sqrt{2}$. The solution for the scattering power Eqs. (\ref{eq:Tsol1}) can be expressed as a function of the reduced momentum transfer.
%\begin{eqnarray}
\begin{equation}
s(q) = \frac{\sigma(q)}{\mu(q)} \frac{e^{-\mu(q)T}}{f(q)}  \Big[ 1 - e^{-\mu(q) T f(q)} \Big], %\\ \nonumber
%f(q) = \frac{2q^2}{1-2q^2}
%s(q) = \frac{\sigma_0(q)}{\mu(q)} \frac{1-2q^2}{2q^2} e^{-\mu(q)T} \Big[1 - e^{-\mu(q) T 2q^2/(1 - 2q^2)} \Big] \\ \nonumber
%T_{opt} = \frac{1}{\mu(q)} \frac{-\ln(1 - 2q^2)(1-2q^2)}{2q^2}; \  % do I really need this?
%s_{opt} = \frac{\sigma_0(q)}{\mu(q)} (1-2q^2)^{\frac{1}{2q^2}}     % and this?
%\exp\bigg[- \frac{\ln b }{b - 1} \bigg]
\label{eq:Tsolq0}
%\end{eqnarray}
\end{equation}

where
\begin{equation}
f(q) = \frac{2q^2}{1-2q^2}
\end{equation}

In the reflection geometry ($1/\sqrt{2} < q \leq 1$) the dependence on the momentum transfer can be derived similarly. 
Assuming normal incidence $\gamma_0$~=~1 and using Eq.~(\ref{eq:q}), Eq.(\ref{eq:Rsol1}) can be written as

\begin{equation}
s(q) = \frac{\sigma(q)}{\mu(q)} \frac{1}{f(q)} \Big[e^{\mu(q) T f(q)} - 1 \Big] 
\label{eq:Rsolq0}
\end{equation}

The leading term in the expressions for the scattering power is the scattering-attenuation contrast $\sigma/\mu$. The scattering geometry defines the measured 
direction-dependent fraction of the scattering-attenuation contrast. If the latter is denoted as $F(q)$ the scattering power is
%Eqs.~(\ref{eq:Tsolq0}) and \ref~{eq:Rsolq0} can be combined and the scattering power can be written as
\begin{eqnarray}
s &=& \frac{\sigma(q)}{\mu(q)}F(q), \\ \nonumber
F(q) &=& 
\begin{cases}
    \frac{e^{-\mu(q)T}}{f(q)} \Big[ 1 - e^{-\mu(q) T f(q)} \Big], & \text{if } 0 \leq q < 1/\sqrt{2},
    \\
     \frac{1}{f(q)} \Big[e^{\mu(q) T f(q)} - 1 \Big], & \text{if } 1/\sqrt{2} < q \leq 1
\end{cases}
\label{eq:sq}
\end{eqnarray}

\begin{figure}
\caption{Geometric factor $F(q)$ (case of normal incidence) for different values of slab thickness ($\mu T$): 
(a) transmission geometry, 0.05 $\leq \mu T\leq $1.0; (b) reflection geometry, 0.05 $\leq \mu T\leq $1.0; (c) transmission geometry, 1.0 $\leq \mu T\leq $4.0; 
(d) reflection geometry, 1.0 $\leq \mu T\leq $4.0. The maximum value of $F(q)$ ($F(0)$ for transmission and $F(1)$ for reflection) is monotonically 
increasing with $\mu T$ for (a) (b) and (d) and decreasing for (c).} 
\includegraphics[scale=0.5]{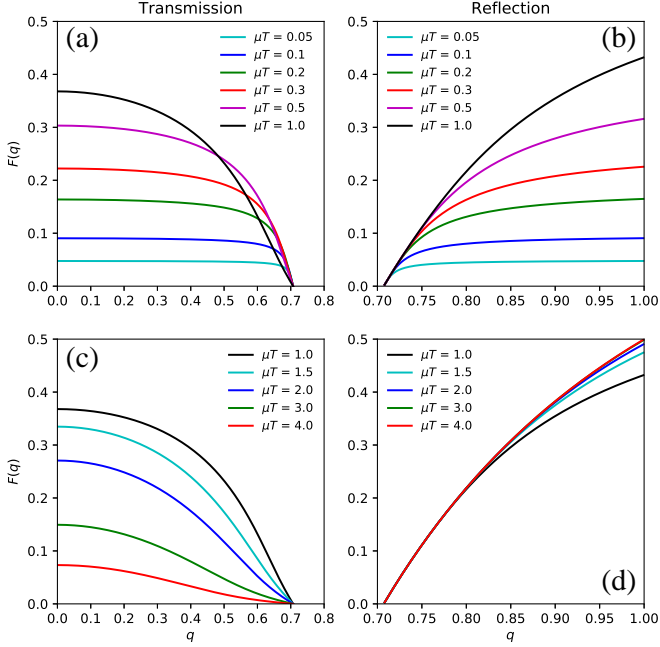} 
\label{fig:Fq0}
\end{figure}

The geometric factor $F(q)$ is plotted in Fig.~\ref{fig:Fq0} for different values of slab thickness in both transmission and reflection geometries. 
The dependence $\mu(q)$ is not considered, which corresponds to a case where total attenuation is dominated by isotropic processes. 
$F(q)$ is the thickness and momentum-transfer-dependent correction to the scattering power. 
The correction is non-negligible in general and of particular importance in the regime of wide angle scattering for $\mu T \gtrsim$~1. 
$F(q)$ normalized to its maximum value exhibits different shapes with variation in $\mu T$ (see Appendix~\ref{ap:0}).

In the transmission geometry the geometric factor $F(q)$ attains its maximum values $e^{-1}$ 
in the regime of small angle scattering ($q \approx$ 0). In the reflection geometry $F(q)$ attains its maximum value 1/2 at backscattering ($q = 1$) for an infinitely thick slab converging to this value rapidly for $\mu T \gtrsim$~2 (see Fig.\ref{fig:Fq0}(d)).

Analysis in the more general case of arbitrary angle of incidence is given in Appendix~\ref{ap:1}. Increase in $F(q)$ beyond the limiting values 
$e^{-1}$ and 1/2 can be obtained by choice of $\alpha$. This can lead to an enhancement in the measured scattered intensity as long as the scattering geometry remains practical and the resulting scattered beam is fully intercepted by the detector. 

%%%%%%%%%%%%%%%%%%%%%%%%%%%%%%%%%%%%%%%%%%%%%%%%%%%%%%%%%%%%%%%%%%%%%%%%%%%%%%%%%%%%%%%%%%%%%%%%%%%%%%%%%%%%%%%%%%%%%%%%%%%%%%
\section{On the scattering-attenuation contrast}
\begin{figure}[!h]
\label{fig:sig-mu}
\caption{Scattering-attenuation contrast $\sigma(E)/\mu(E)$ using XCOM database for selected compounds. Solid lines correspond to coherent (elastic) scattering and dashed lines correspond to incoherent (inelastic) scattering.} 
\includegraphics[scale=0.5]{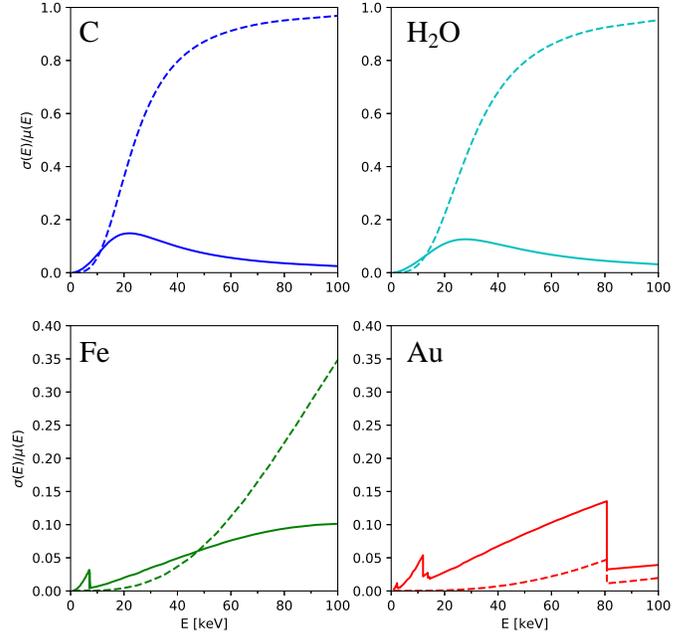} 
\end{figure}

% plot sigma/mu*e-1 (opt. transmission) and sigma/mu*1/2 backreflection for several elements!
%Studies of weak scattering signals benefit from the optimal choice of sample thickness at a given photon energy.
For the special cases discussed above, small angle scattering and backscattering, 
the variation in the geometric factor with $q$ can be neglected. 
The optimal scattering power is defined only by the scattering-attenuation contrast.  
It is informative to explore dependence of the scattering-attenuation contrast on the photon energy. 
The contrast is plotted in Fig.\ref{fig:sig-mu} for several representative compounds using values of the scattering and attenuation cross sections taken from XCOM database \cite{nist_xcom}. The solid curves represent coherent (elastic) scattering and the dashed curves represent incoherent (inelastic) scattering. For illustrative purposes it is assumed that the scattering is isotropic. The focus is on comparison of fractions of coherently and incoherently scattered intensity, which gives an insight on the optimal choice of photon energy in non-resonant X-ray scattering experiments. 

For compounds of low atomic numbers such as carbon and water the elastic component dominates up to about 10-15 keV (as shown in Fig.\ref{fig:sig-mu}). Above these popular energies the enhancement in inelastic scattering  is rather significant, which makes selective detection of elastic scattering more difficult. If, however, inelastic scattering is suppressed using energy discrimination, the elastic component is maximized at photon energies of about 20-30 keV.
For high-Z compounds (Fe and Au shown in Fig.\ref{fig:sig-mu}) the scattering-attenuation contrasts are monotonically increasing with photon energy (aside from transitions due to the photoabsorption edges). For Fe the inelastic component begins to dominate at $\approx$~50~keV  while for Au the elastic component remains dominant up to 100 keV and beyond. 
It is clear that for all compounds inelastic scattering experiments should benefit from the choice of higher photon energies due to enhancement in the corresponding scattering-attenuation contrast (shown by the dashed lines in Fig.\ref{fig:sig-mu}). 
%At high energies ($E \gtrsim $~30~keV) X-ray crystal analyzers, which provide high-resolution energy discrimination gradually become impractical due to a reduction of the intrinsic angular and energy acceptances of Bragg reflections. Nevertheless, further development of high-efficiency nonresonant inelastic X-ray spectrometers operating at greater photon energies can be justified by the rapid increase in the inelastic scattering coefficient for low-Z materials. 

It should be noted that in the practical range of photon energies considered here (up to 100~keV) the scattering-attenuation contrast for elastic  X-ray scattering remains small for either low-Z or high-Z compounds ($\sigma/\mu \lesssim 0.2$). On the contrary, the scattering-attenuation contrast for inelastic X-ray scattering from low-Z materials gradually approaches unity at higher photon energies, thus violating the approximation of negligible multiple scattering.  
Finally, it is noted that practical calculation of scattered intensity measured in an experiment should include considerations of the scattering cross section of interest (from which $\sigma(q)$ is derived), polarization effects and detector/analyzer acceptances.

%%%%%%%%%%%%%%%%%%%%%%%%%%%%%%%%%%%%%%%%%%%%%%%%%%%%%%%%%%%%%%%%%%%%%%%%%%%%%%%%%%%%%%%%%%%%%%%%%%%%%%%%%%%%%%%%%%%%%%%%%%%%%%%%%%%%%%%%%%%%%%%%%%%%%%%%
\section{Conclusions}
%%%%%%%%%%%%%%%%%%%%%%%%%%%%%%%%%%%%%%%%%%%%%%%%%%%%%%%%%%%%%%%%%%%%%%%%%%%%%%%%%%%%%%%%%%%%%%%%%%%%%%%%%%%%%%%%%%%%%%%%%%%%%%%%%%%%%%%%%%%%%%%%%%%%%%%%
In conclusion, X-ray scattering power of a plane parallel homogeneous slab of material was derived using radiation transfer considerations. 
Arbitrary angles of incidence and emergence were considered using an analogy with reflection geometry of an asymmetric crystal.  
%Effects of total external reflection in grazing incidence or emergence were not considered. 
In the case of small angle scattering in transmission through the slab the expressions are reduced to the well known 
"rule of thumb" where the optimal sample thickness is inversely proportional to the total attenuation coefficient. 
In the general case, the parallel slab geometry yields direction-dependent corrections to scattered intensity. These corrections are essential in the regime of wide-angle scattering for samples with thickness approximately equal to or greater than the optimal thickness.
The correction was formulated as momentum transfer-dependent geometric factor. The scattering power is described as the product of the photon energy-dependent scattering-attenuation contrast and the derived geometric factor. The dependence of the scattering-attenuation contrast on the photon energy was explored for both elastic and inelastic scattering, assuming isotropic processes for simplicity. The results can be used in finding optimal experimental conditions to enhance weak scattering (both elastic and inelastic) of interest for samples with plane parallel slab geometry.
%%%%%%%%%%%%%%%%%%%%%%%%%%%%%%%%%%%%%%%%%%%%%%%%%%%%%%%%%%%%%%%%%%%%%%%%%%%%%%%%%%%%%%%%%%%%%%%%%%%%%%%%%%%%%%%%%%%%%%%%%%%%%%%%%%%%%%%%%%%%%%%%%%%%%%%%%%%%%%%%%%%%%%%%%%%%%%
\acknowledgements
J.P.C. Ruff, D.M. Smilgies and C. Franck are acknowledged for helpful discussions.
This work is based upon research conducted at the Cornell High Energy Synchrotron Source (CHESS) 
which is supported by the National Science Foundation under award DMR-1332208.
%%%%%%%%%%%%%%%%%%%%%%%%%%%%%%%%%%%%%%%%%%%%%%%%%%%%%%%%%%%%%%%%%%%%%%%%%%%%%%%%%%%%%%%%%%%%%%%%%%%%%%%%%%%%%%%%%%%%%%%%

\appendix
\section{Case of normal incidence ($\gamma_0 = 1$): \newline Normalized geometric factor $F(q)$.}\label{ap:0}

\begin{figure}[h!]
\includegraphics[width=0.5\textwidth]{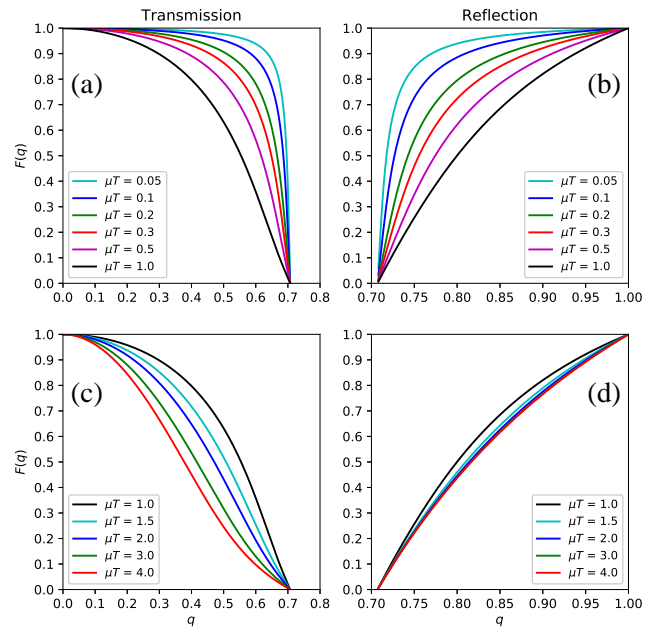}
\caption{Geometric factor $F(q)$ normalized by its maximum value (normal incidence) plotted for different values of slab thickness ($\mu T$): 
(a) transmission geometry, 0.05 $\leq \mu T\leq $1.0; (b) reflection geometry, 0.05 $\leq \mu T\leq $1.0; (c) transmission geometry, 1.0 $\leq \mu T\leq $4.0; 
(d) reflection geometry, 1.0 $\leq \mu T\leq $4.0. The maximum value of $F(q)$ ($F(0)$ for transmission and $F(1)$ for reflection) is monotonically 
increasing with $\mu T$ for (a) (b) and (d) and decreasing for (c).}
\label{fig:gfunc2n}
\end{figure}

\section{Dependence on the momentum transfer for X-rays: arbitrary angle of incidence}\label{ap:1}

For the more general case of arbitrary angle of incidence the geometric factor $F(q)$ can be obtained as follows. The asymmetry ratio $b = \gamma_0/\gamma_s$ can be expressed in terms of incidence angle $\alpha$ 
($\gamma_0 = \cos\alpha$) and the scattering angle $\theta$:
\begin{equation}
b = \frac{\cos\alpha}{\cos(\theta \pm \alpha)}
\label{eq:b}
\end{equation}

\begin{figure}
\includegraphics[width=0.5\textwidth]{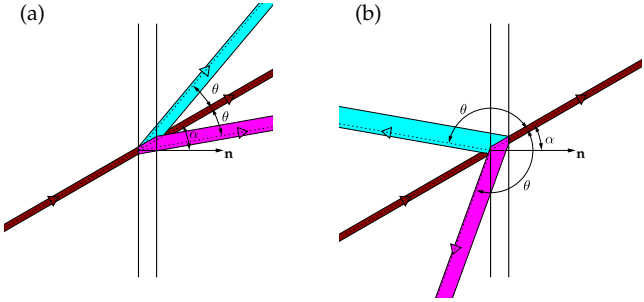}
\caption{Two possible scattering branches shown in (a) transmission geometry and (b) reflection geometry.
The incident beam is shown with dark red color.}
\label{fig:br}
\end{figure}

The $\pm$ sign defines two possible scattering branches. They are shown in Fig.~\ref{fig:br}(a) and \ref{fig:br}(b) for the transmission and reflection geometry, respectively. The beam marked by the cyan color corresponds to the upper ("+") sign in Eq.~(\ref{eq:b}) and the beam marked by the magenta color corresponds to the lower sign. 

Using the reduced momentum transfer $q = \frac{Q}{2k_0} = \sin(\theta/2)$ expansion of the denominator of Eq.(\ref{eq:b}) yields:

\begin{equation}
\gamma_s = (1 - 2q^2)\cos\alpha \mp \sqrt{1 - (1-2q^2)^2} \sin\alpha
\end{equation}

The expression for the geometric factor $F(q)$ given by Eqs.~(23) remains valid if $\mu T$ is replaced with $\mu T/\gamma_0$ and the function $f(q)$ becomes:
\begin{equation}
f(q) = \frac{2q(q \pm \sqrt{1-q^2})\tan\alpha}{1-2q^2 \mp 2q\sqrt{1-q^2}\tan\alpha}
\end{equation}

\begin{figure}[h!]
\includegraphics[width=0.5\textwidth]{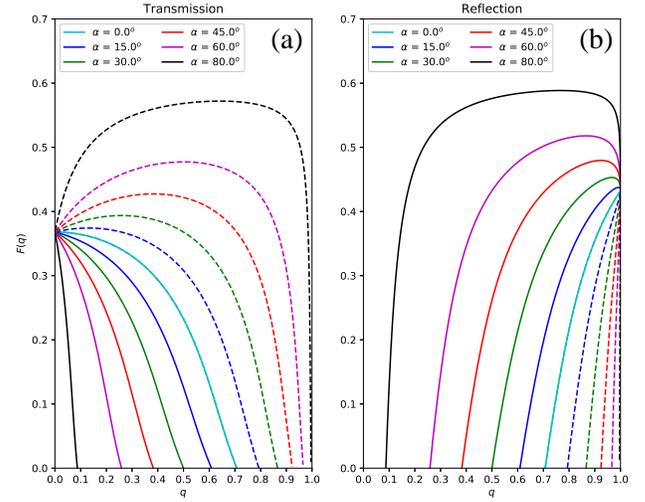}
\caption{Geometric factor $F(q)$ for $\mu T/\gamma_0 = 1$ plotted using several different values of the incidence angle $ 0^o \leq \alpha < 90^o$: (a) transmission scattering geometry, and (b) reflection scattering geometry. The geometric factors corresponding to the two distinct scattering branches are shown with either solid or dashed lines.}
\label{fig:qfunc1}
\end{figure}

For the case $\mu T/\gamma_0 = 1$ the geometric factor for the transmission and reflection geometries is shown in Fig.\ref{fig:qfunc1} for several different values of the incidence angle $\alpha$ (color coded as shown in the Figure legend). One of the scattering branches is shown with solid lines and the other - with dashed lines. Obviously, the two merge for
$\alpha = 0^o$.

%%%%%%%%%%%%%%%%%%%%%%%%%%%%%%%%%%%%%%%%%%%%%%%%%%%%%%%%%%%%%%%%%%%%%%%%%%%%%%%%%%%%%%%%%%%%%%%%%%%%%%%%%%%%%%%%%%%%%%%%%%%%%%%%%%%%%%%%%%%%%%%%%%%%%%%%%%%%%%%%%%%%%%%%%%%%%

%\bibliography{/nfs/chess/aux/user/sas456/Work3/Literature/bib/references}
%\bibliography{/home/sstoupin/Work3/Literature/bib/references}

%%%%%%%%%%%%%%%%%%%%%%%%%%%%%%%%%%%%%%%%%%%%%%%%%%%%%%%%%%%%%%%%%%%%%%%%%%%%%%%%%%%%%%%%%%%%%%%%%%%%%%%
\end{document}